# Simulating Self-gravitating Hydrodynamic Flows

EWALD MÜLLER and MATTHIAS STEINMETZ
MAX-PLANCK-INSTITUT FÜR ASTROPHYSIK,
KARL-SCHWARZSCHILD-STR. 1, 85740 GARCHING, FRG

An efficient algorithm for solving Poisson's equation in two and three spatial dimensions is discussed. The algorithm, which is described in detail, is based on the integral form of Poisson's equation and utilizes spherical coordinates and an expansion into spherical harmonics. The solver can be applied to and works well for all problems for which the use of spherical coordinates is appropriate. We also briefly discuss the implementation of the algorithm into hydrodynamic codes which are based on a conservative finite–difference scheme.

## 1. Introduction

Self-gravity often plays a major role when simulating hydrodynamic flows in astrophysics. While it is trivial to include the effects of self-gravity in one-dimensional (Lagrangian) hydrodynamical simulations, the problem becomes much more difficult for multi-dimensional problems requiring the numerical solution of Poisson's equation in two or three spatial dimensions in each hydrodynamical time step. During the last three decades various algorithms have been proposed to solve this task efficiently on a computer. These algorithms can be divided into two major categories which solve Poisson's equation in differential or integral form, respectively.

The first category of Poisson solvers consists of several distinct sub-categories of algorithms. A common feature of all these algorithms is that one discretizes the differential form of Poisson's equation

$$\Delta \Phi = 4\pi G \varrho(\mathbf{r}) \qquad (1)$$

on a grid of $N = N_1 \times N_2 \times N_3$ zones. Here $\Phi$, $G$, $\varrho$ and $N_i$ are the gravitational potential, the gravitational constant, the density, and the number of zones in the $i$-th coordinate direction of the three–dimensional grid, respectively. In its simplest form, i.e., using a local seven–point (five–point in 2D) finite–difference approximation to the Laplace operator $\Delta$, the discretization process gives rise to a sparse matrix problem (see e.g., Potter 1973). The sparse matrix is of block hepta–diagonal (penta–diagonal in 2D) form where the diagonal block is itself tri–diagonal and the off–diagonal blocks are diagonal. This specific form of the matrix also holds for non-uniform grids as long as the grid is topologically rectangular and each vertex has eight (or four in 2D) adjacent zones. Because of its simple pattern and because most of the blocks are entirely zero the operation count for inverting the matrix reduces to $O(N^2)$ instead of the $O(N^3)$ required for an arbitrary matrix of dimension $N$. However, one would like algorithms for inverting the matrix arising from the multi–dimensional Laplace operator in $O(N \ln N)$ operations, because the inversion has to be done once in each hydrodynamical timestep. Several such algorithms can be found in the literature. One group of algorithms is based on Fourier analysis techniques (Hockney 1970, Boris & Roberts 1969) which allows one to exploit the standard, highly optimized Fast Fourier Transform (FFT) algorithms, and on a cyclic reduction (or folding) technique along spatial dimension (Buneman 1969). Both algorithm, however, do not allow (or at least require an additional interpolation) for non-uniform zones, an irregularly shaped grid, or complicated boundary conditions.

These restrictions do not hold for the second sub–category of finite–difference algorithm which are based on iterative techniques such as Jacobi iteration, Gauss–Seidel iteration, Successive Over–Relaxation (SOR), and cyclic Chebyshev Method (for details see, e.g., Potter 1973; Oran & Boris 1987). These methods are easy to program, but are computationally expensive relative to faster direct solution methods. A somewhat more complex iterative method is called Alternating–Direction Implicit Method (ADI), which involves successive iterated implicit steps in each spatial direction, where each step requires the solution of a tri–diagonal linear system. All these iterative methods use the structure of the grid and the sparsity of the matrix to reduce the operation count. As each iteration requires of order $N$ operations, and as at least $\max(N_1, N_2, N_3)$ iterations have to be performed to "propagate" information over the entire grid, the operation count for obtaining a solution is $O(N^{3/2})$ in two dimensions and $O(N^{4/3})$ in three dimensions (see e.g., Oran & Boris 1987). A further iterative method is called Incomplete Cholesky Conjugate Gradient (ICCG). It was proposed by Meijerink & van der Vorst (1977) and by Kershaw (1978) and solves the discretized Poisson equation (or any other symmetric linear problem) by a Conjugate Gradient (CG) iteration with some kind of "pre–conditioning" of the corresponding matrix. With the latter procedure one achieves a significantly faster convergence of the CG iteration, which without "pre–conditioning" requires $N$ iterations to obtain a solution. Finally, one can use the multigrid iterative approach (Federenko 1961; Brandt 1977, 1981; Douglas 1984) to solve the discretized Poisson equation combining direct and iterative methods and thus lowering the operation count to $O(N \ln N)$ (for more details see, e.g., the book edited by Hackbusch & Trottenberg 1983).

The second category of algorithms solve Poisson's equation in integral form. Using the theory of Greens functions the formal solution of the Poisson equation (Eq. (1)) can be written in the form

$$\Phi = -G \int d^3\mathbf{r}' \, \frac{\varrho(\mathbf{r}')}{|\mathbf{r} - \mathbf{r}'|}, \qquad (2)$$

provided the potential and its gradient vanish for $r \to \infty$. The advantage of the integral form is twofold. Firstly the boundaries of the grid cause no problems and do not require any special handling, and secondly the potential of arbitrarily strongly varying density distributions including even point masses can be computed.

In principle, it would be possible to integrate Eq. (2) numerically on a grid, where the integration could be performed either in Fourier space by FFT techniques or in ordinary space by standard integration formulas, like e.g., by the second-order accurate trapezoidal rule. In the case of a star like object or of a not too flattened ellipsoidal matter distribution, spherical coordinates $(r, \theta, \phi)$ would be the appropriate choice to perform such an integration. However, in order to obtain the potential at one location a sum over $N \equiv N_r \times N_\theta \times N_\phi$ grid cells has to be computed, where $N_r$, $N_\theta$ and $N_\phi$ are the number of zones in $r$, $\theta$ and $\phi$ direction, respectively. Thus, computing the potential for all grid cells implies an operation count proportional to $\frac{1}{2} N (N+1) \approx N^2$. This becomes prohibitively expensive even for moderate size grids consisting of a few thousands zones (i.e., a few tens per dimension in 3D).

In order to reduce the computational cost essentially to a $\propto N$ behaviour the volume integral in Eq. (2) is approximated by a triple (or double in 2D) summation and the denominator in the Poisson integral is expanded in terms of spherical harmonics. The expansion allows one to separate the problem in radial and angular direction, i.e., the multi–dimensional problem is reduced into several one–dimensional ones. The potential of each zone is obtained by calculating the moments of the mass distribution inside and outside a sphere defined by the radius of that zone. The computational costs of the resulting algorithm scales as $(L+1)^2 \times N_r \times N$, where $L$ is the order of the highest spherical harmonics used in the expansion. Exploiting analytical properties of the moment integrals the algorithm can be made more efficient the computational costs being proportional to $L^2 \times N$, only (Tschaepe 1987, Binney & Tremaine 1987).

## 2. An efficient Poisson solver

### 2.1 The algorithm

The integrand of Eq. (2) can be expanded into spherical harmonics $Y^{lm}(\theta, \phi)$, where the superscripts $l$ and $m$ uniquely identify each spherical harmonic. One then obtains

$$\frac{1}{|\mathbf{r} - \mathbf{r}'|} = \sum_{l=0}^{\infty} \sum_{m=-l}^{l} \frac{4\pi}{2l+1} Y^{lm}(\theta, \phi) Y^{lm*}(\theta', \phi') \frac{r_<^l}{r_>^{l+1}}, \qquad (3)$$

where $r_<$ ($r_>$) is the smaller (larger) of the two radii $r$ and $r'$, and where $Y^{lm*}$ denotes the complex conjugate of $Y^{lm}$. Consequently, the gravitational potential is given by

$$\Phi(r, \theta, \phi) = -G \sum_{l=0}^{\infty} \frac{4\pi}{2l+1} \sum_{m=-l}^{l} Y^{lm}(\theta, \phi) \int dr' \, r'^2 \, d\Omega' \, \varrho(r', \theta', \phi') \frac{r_<^l}{r_>^{l+1}} Y^{lm*}(\theta', \phi') \qquad (4)$$

$$= -G \sum_{l=0}^{\infty} \frac{4\pi}{2l+1} \sum_{m=-l}^{l} Y^{lm}(\theta, \phi) \left( \frac{1}{r^{l+1}} C^{lm}(r) + r^l D^{lm}(r) \right) \qquad (5)$$

with

$$C^{lm}(r) = \int_{4\pi} d\Omega' \, Y^{lm*}(\theta', \phi') \int_0^r dr' \, r'^{l+2} \, \varrho(r', \theta', \phi'), \qquad (6)$$

$$D^{lm}(r) = \int_{4\pi} d\Omega' \, Y^{lm*}(\theta', \phi') \int_r^{\infty} dr' \, r'^{1-l} \, \varrho(r', \theta', \phi'), \qquad (7)$$

where $d\Omega \equiv \sin\theta \, d\theta \, d\phi$. Note that the functions $C^{lm}(r)$ and $D^{lm}(r)$ do not depend on the angular coordinates $(\theta, \phi)$ of the point where the potential is to be calculated, because the angular dependence of the potential is completely determined by the spherical harmonics (see Eqs. (4) and (5)).

Up to now no approximation has been made, i.e., Eq. (5) is an exact solution of the Poisson equation. However, this no longer holds when one truncates the sum in Eq. (5) at $l = L$, i.e., when omits all spherical harmonics with $l > L$. The truncation error resulting from this approximation obviously is problem-dependent. Thus, depending on the density distribution different values of $L$ must be chosen to reduce the error below a certain accuracy. In practice, however, a fixed value of $L \approx 12$ can be used for quite different matter distributions of spheroidal or ellipsoidal shape without introducing errors larger than a few percent (see Fig. 2).

Since there exist $2L + 1$ spherical harmonics of order $L$ and consequently $(L+1)^2$ spherical harmonics of order smaller than or equal to $L$, the computational effort of solving the Poisson equation can be reduced from an operation count $\propto N_r^2 \times N_\theta^2 \times N_\phi^2$ to one $\propto (L+1)^2 N_r^2 \times$ $N_\theta \times N_\phi$ when using Eq. (5) with the first sum truncated at $l = L$. Obviously, the truncated series expansion into spherical harmonics becomes a very useful approximation, if the density distribution is almost spherically symmetric, like e.g., in a slowly rotating star. In that case even with a few spherical harmonics very accurate results can efficiently be obtained the operation count $\propto N_\theta \times N_\phi$ being reduced to a much smaller one $\propto (L+1)^2$. In the case of a highly flattened structure, however, a large number of spherical harmonics must be taken into account, i.e., the effort to calculate the corresponding $(L+1)^2$ moments becomes comparable to that of calculating the double sum over $N_\theta$ and $N_\phi$.

A further optimization of the operation count can be achieved when the discretized implementation of the proposed Poisson solver is considered. We split all integrals in Eq. (5) into a sum of integrals over subintervals corresponding to the respective boundaries of the cells of the three-dimensional computational grid. Then the potential at position $(r_n, \theta_o, \phi_p)$ is given by

$$\Phi(r_n, \theta_o, \phi_p) = -G \sum_{l=0}^{L} \frac{4\pi}{2l+1} \sum_{m=-l}^{l} Y^{lm}(\theta_o, \phi_p) \left( \frac{1}{r_n^{l+1}} C_n^{lm} + r_n^l D_n^{lm} \right) \tag{8}$$

with

$$C_n^{lm} = \sum_{j=1}^{N_\theta} \sum_{k=1}^{N_\phi} \left\{ \int_{\phi_{k-1}}^{\phi_k} \int_{\theta_{j-1}}^{\theta_j} \sin\theta \, d\theta \, d\phi \, Y^{lm*}(\theta, \phi) \sum_{i=1}^{n} \int_{r_{i-1}}^{r_i} dr \, r^{l+2} \, \varrho(r, \theta, \phi) \right\} \tag{9}$$

$$D_n^{lm} = \sum_{j=1}^{N_\theta} \sum_{k=1}^{N_\phi} \left\{ \int_{\phi_{k-1}}^{\phi_k} \int_{\theta_{j-1}}^{\theta_j} \sin\theta \, d\theta \, d\phi \, Y^{lm*}(\theta, \phi) \sum_{i=n+1}^{N_r} \int_{r_{i-1}}^{r_i} dr \, r^{1-l} \, \varrho(r, \theta, \phi) \right\} \tag{10}$$

where the subscripts $n, o$ and $p$ are running from 1 to $N_r$, $N_\theta$ and $N_\phi$, respectively

From Eqs. (9) and (10) it is obvious that a computational effort $\propto N_r$ is necessary to calculate one of the coefficients $C_n^{lm}$ or $D_n^{lm}$, i.e., the total computational cost depends quadratically on $N_r$. As the number of radial grid points $N_r$ is larger than $N_\theta$ and $N_\phi$ in most simulations, the quadratic dependence on $N_r$ is unwanted. However, the operation count can be reduced to one $\propto N_r$ by exploiting the following recursion relations of the radial integrals:

$$\int_0^{r_{i+1}} dr' \, r'^{l+2} \, \varrho(r', \theta', \phi') = \int_0^{r_i} dr' \, r'^{l+2} \, \varrho(r', \theta', \phi') + \int_{r_i}^{r_{i+1}} dr' \, r'^{l+2} \, \varrho(r', \theta', \phi') \tag{11}$$

$$\int_{r_{i-1}}^{r_{N_r}} dr' \, r'^{1-l} \, \varrho(r', \theta', \phi'), = \int_{r_i}^{r_{N_r}} dr' \, r'^{1-l} \, \varrho(r', \theta', \phi') + \int_{r_{i-1}}^{r_i} dr' \, r'^{1-l} \, \varrho(r', \theta', \phi'). \tag{12}$$

Hence, all coefficients $C_n^{lm}$ and $D_n^{lm}$ can be calculated with an operation count $\propto (L+1)^2 N_r \times N_\theta \times N_\phi$.

The proposed algorithm to compute the gravitational potential of an arbitrary three-dimensional mass distribution can now be summarized.

*Step 1*: Calculate the angular and radial weights.

$$A_{ijk}^{lm} = \int_{\phi_{k-1}}^{\phi_k} \int_{\theta_{j-1}}^{\theta_j} \int_{r_{i-1}}^{r_i} \sin\theta \, d\theta \, d\phi \, dr \, r^{l+2} \, Y^{lm*}(\theta, \phi) \, \varrho(r, \theta, \phi) \tag{13}$$

$$B_{ijk}^{lm} = \int_{\phi_{k-1}}^{\phi_k} \int_{\theta_{j-1}}^{\theta_j} \int_{r_{i-1}}^{r_i} \sin\theta \, d\theta \, d\phi \, dr \, r^{1-l} \, Y^{lm*}(\theta, \phi) \, \varrho(r, \theta, \phi) \tag{14}$$

*Step 2*: Calculate the $(l,m)$-th moment of the inner and outer mass distribution using the recursion formula given in Eqs. (11) and (12).

$$C_n^{lm} = \sum_{i=1}^{n}\sum_{j=1}^{N_\theta}\sum_{k=1}^{N_\phi} A_{ijk}^{lm} = C_{n-1}^{lm} + \sum_{j=1}^{N_\theta}\sum_{k=1}^{N_\phi} A_{njk}^{lm} \qquad (15)$$

$$D_n^{lm} = \sum_{i=n}^{N_r}\sum_{j=1}^{N_\theta}\sum_{k=1}^{N_\phi} B_{ijk}^{lm} = D_{n+1}^{lm} + \sum_{j=1}^{N_\theta}\sum_{k=1}^{N_\phi} B_{njk}^{lm} \qquad (16)$$

*Step 3*: Compute the potential by summing up the individual contributions.

$$\Phi(r_n, \theta_o, \phi_p) = -G \sum_{l=0}^{L} \frac{4\pi}{2l+1} \left\{ \frac{1}{(r_n)^{l+1}} \left[ \sum_{m=-l}^{l} Y^{lm}(\theta_o, \phi_p)\, C_n^{lm} \right] \right. \qquad (17)$$

$$\left. + (r_n)^l \left[ \sum_{m=-l}^{l} Y^{lm}(\theta_o, \phi_p)\, D_n^{lm} \right] \right\} \qquad (18)$$

## 2.2 Coding, Accuracy and Performance

As shown above, the computational cost of the potential solver grows $\propto (L+1)^2 \times N_r \times N_\theta \times N_\phi$, the most time consuming part being the calculation of the weights $A_{ijk}^{lm}$ and $B_{ijk}^{lm}$, and of the inner and outer moments $C_i^{lm}$ and $D_i^{lm}$. Note, that in general, the calculation of one weight $A_{ijk}^{lm}$ would require a numerical integration! Further note that the weights are time dependent (because of their density dependence), that all weights can be computed independently of each other, and that the weights for the same zone $(i,j,k)$ but for different $(l,m)$ are related to each other via the recursion formula of the spherical harmonics.

These properties can be exploited in the practical implementation of the Poisson solver, which has been done in the FORTRAN version of the 2D solver included on the floppy disk delivered with this volume. In particular, the following points have been considered:

- If the density is a spatially slowly varying function, one can rewrite Eqs. (13) and (14) in the form

$$A_{ijk}^{lm} = \varrho_{ijk} \left( \int_{\phi_{k-1}}^{\phi_k} \int_{\theta_{j-1}}^{\theta_j} \sin\theta\, d\theta\, d\phi\, Y^{lm*}(\theta,\phi) \right) \left( \int_{r_{i-1}}^{r_i} dr\, r^{l+2} \right) \qquad (19)$$

$$B_{ijk}^{lm} = \varrho_{ijk} \left( \int_{\phi_{k-1}}^{\phi_k} \int_{\theta_{j-1}}^{\theta_j} \sin\theta\, d\theta\, d\phi\, Y^{lm*}(\theta,\phi) \right) \left( \int_{r_{i-1}}^{r_i} dr\, r^{1-l} \right) . \qquad (20)$$

These expression are exact for the monopole term, if $\varrho_{ijk}$ is interpreted a zone average. Because this holds for most modern finite difference schemes, and because the monopole contribution is the dominant part for most practical cases, the approximations given in Eqs. (19) and (20) are usually accurately enough the error being of second-order, only.

- The spherical harmonics $Y^{lm}$ can be evaluated very efficiently, if recursion relations are used. Thus, the coefficients $A_{ijk}^{lm}$ and $B_{ijk}^{lm}$ given in Eqs. (13) and (14) can also be determined recursively by integrating the recursion relation. In practice, one first computes all spherical harmonics $Y_{jk}^{lm} = Y^{lm}(\theta_j, \phi_k)$ together with the corresponding angular integrals (see Eqs. (19) and (20)) and stores these quantities for later usage. In a simulation, where the angular zones are time independent, this has to be done only once at the beginning of the calculation.

- In order to reduce the extremely large storage requirements implied by a straightforward implementation of the proposed Poisson solver (because the angular weights would require two five-dimensional arrays, and the moments two three-dimensional ones), one better proceeds as as follows: An outermost loop extends over the indices $(l, m)$ of the spherical harmonics. For each index pair $l, m$ the radial moments $C_n^{lm}$ and $D_n^{lm}$ (see Eqs. (15) and (16)) are calculated using the recursion relation, i.e., the innermost loop runs over the indices $j, k$ of the angular grid. Therefore, the angular weights $A_{ijk}^{lm}$ and $B_{ijk}^{lm}$ (see Eqs. (19) and (20)) need not to be stored in FORTRAN arrays, but can be put into temporary scalars, which are then directly summed up to yield the moments $C_n^{lm}$ and $D_n^{lm}$. To avoid cancellation effects due to summing up terms of nearly identical absolut value but opposite sign, the inner moment $C_n^{lm}$ should be calculated starting from the center of the grid and proceeding outward, while the outer moment $D_n^{lm}$ should be evaluated just the opposite way, i.e., one begins the summation at the border of the grid and the proceeds inwards. Since the moments $C_n^{lm}$ and $D_n^{lm}$ both depend on their respective lower order moments only via spherical harmonics (all of which have already been computed and stored), it is not necessary to keep all moments $C_n^{lm}$ and $D_n^{lm}$ in the central memory, but it is sufficient to store them into temporary vectors $\tilde{C}_n$ and $\tilde{D}_n$. Then the contribution of the $(l, m)$-th moment to the potential $\Phi_{nop}$ can be calculated by multiplying both $\tilde{C}_n$ and $\tilde{D}_n$ with $Y_{op}^{lm}$ and add this contribution to the potential $\Phi_{nop}$ for every point $(r_n, \theta_o, \phi_p)$ for which the potential has to be calculated.

Together with this article, a floppy disk is supplied which contains the source code of the 2D version of the proposed Poisson solver, a test program and subroutines, which provide the analytical solutions for several axisymmetric, self–gravitating configurations: a homogeneous spheroid, a Kuzmin–Plummer, a Satoh and a logarithmic potential. For a definition and a discussion of these configurations we refer to Binney & Tremaine (1987). The shape of the density and potential distribution of these configurations is illustrated in Fig. 1. Note that for all examples, except for the spheroid, the density distribution extends to infinity, i.e., not all mass is within the borders of the (finite) computational grid. Thus, all error estimates given below are obtained by comparing analytical and numerical potential values normalized to their respective central values. In the case of the homeoids we only consider the interior solution for the error estimate.

Because for a homeoid the potential near the center is very smooth, the normalized potential in the first radial zone differs only by $10^{-6}$ from its value at the center. This small difference can become comparable to the rounding error encountered when calculating a homeoid's dipole moment without explicitly assuming equatorial symmetry. Consequently, the comparison of the analytical and numerical solution would erroneously indicate a large error in the potential. Thus, we have eliminated the dipole terms in the poisson solver routine, which implies no restriction as long as the center of the grid coincides with the center of mass of the configuration. Of course, the dipole terms must be included, whenever the center of mass is offset with respect to the computational grid.

Two further questions must still be addressed: When can the proposed Poisson solver be applied and how many multipole moments have to be taken into account. Obviously, the geometry of the spherical grid used for the solver already implies, that the algorithm is well suited for spheroidal configurations. For disk-like configurations the applicability of the method becomes worse with increasing flatness, firstly because more spherical harmonics have to be taken into account, and secondly because a smaller part of the (spherical) grid is occupied by the configuration (see e.g., the highly flattened ellipsoid in Fig. 1). The proposed method also becomes less appropriate, if the problem to be solved involves two or more spatially separated distinct mass concentrations, as it is for example the case for a binary system. Such a case, however,

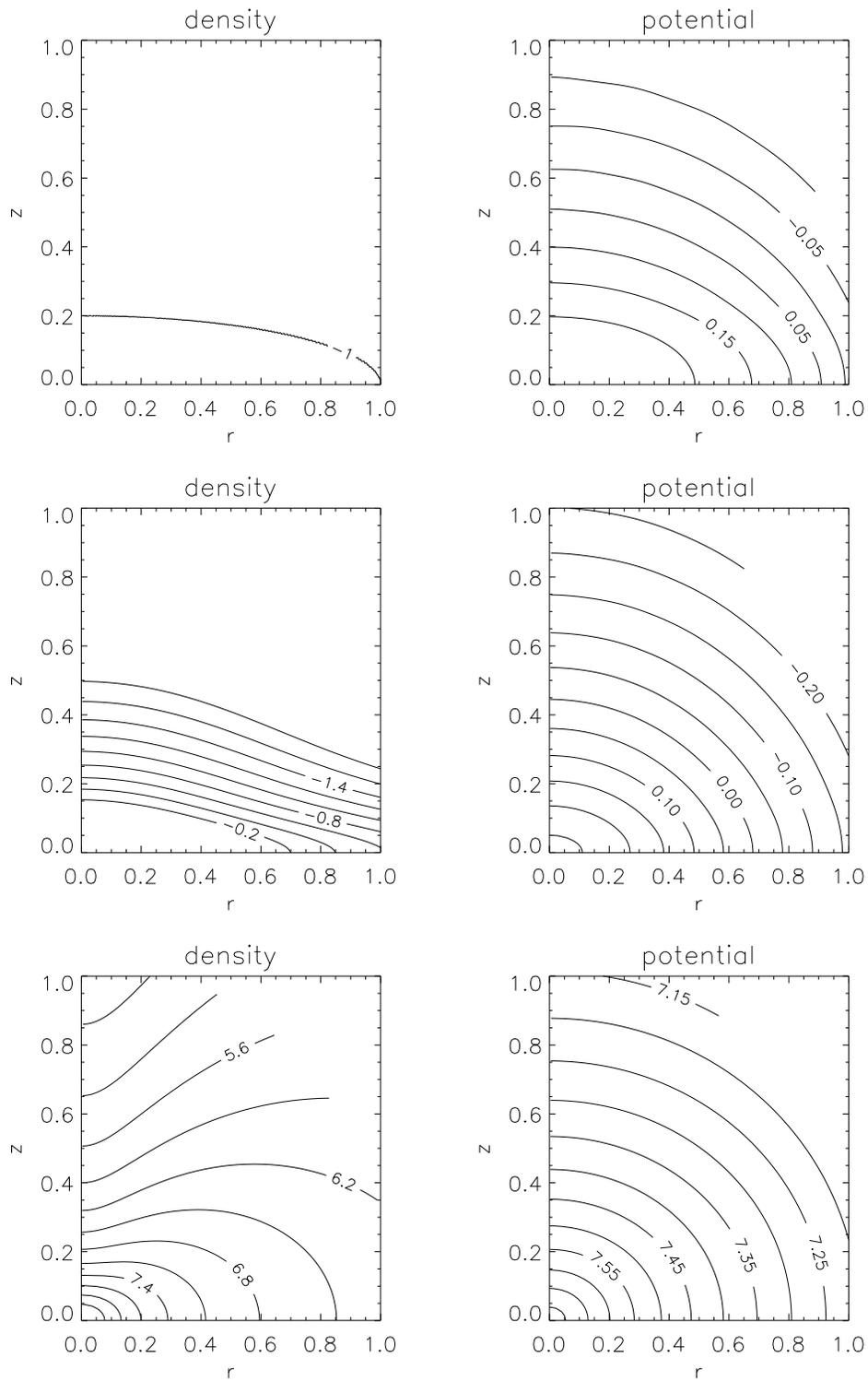

Figure 1: Density (left) and potential (right) of different axisymmetric matter distributions, which all have analytical solution for the potential. Shown are from top to bottom: A homogeneous spheroid with an axis ratio of 0.2, a Kuzmin-Plummer disk with parameters $a = 0.4$ and $b = 0.1$, and a logarithmic potential with $r_c = 0.1$ and $q = 0.52$ ( see Binney & Tremaine 1987)
.

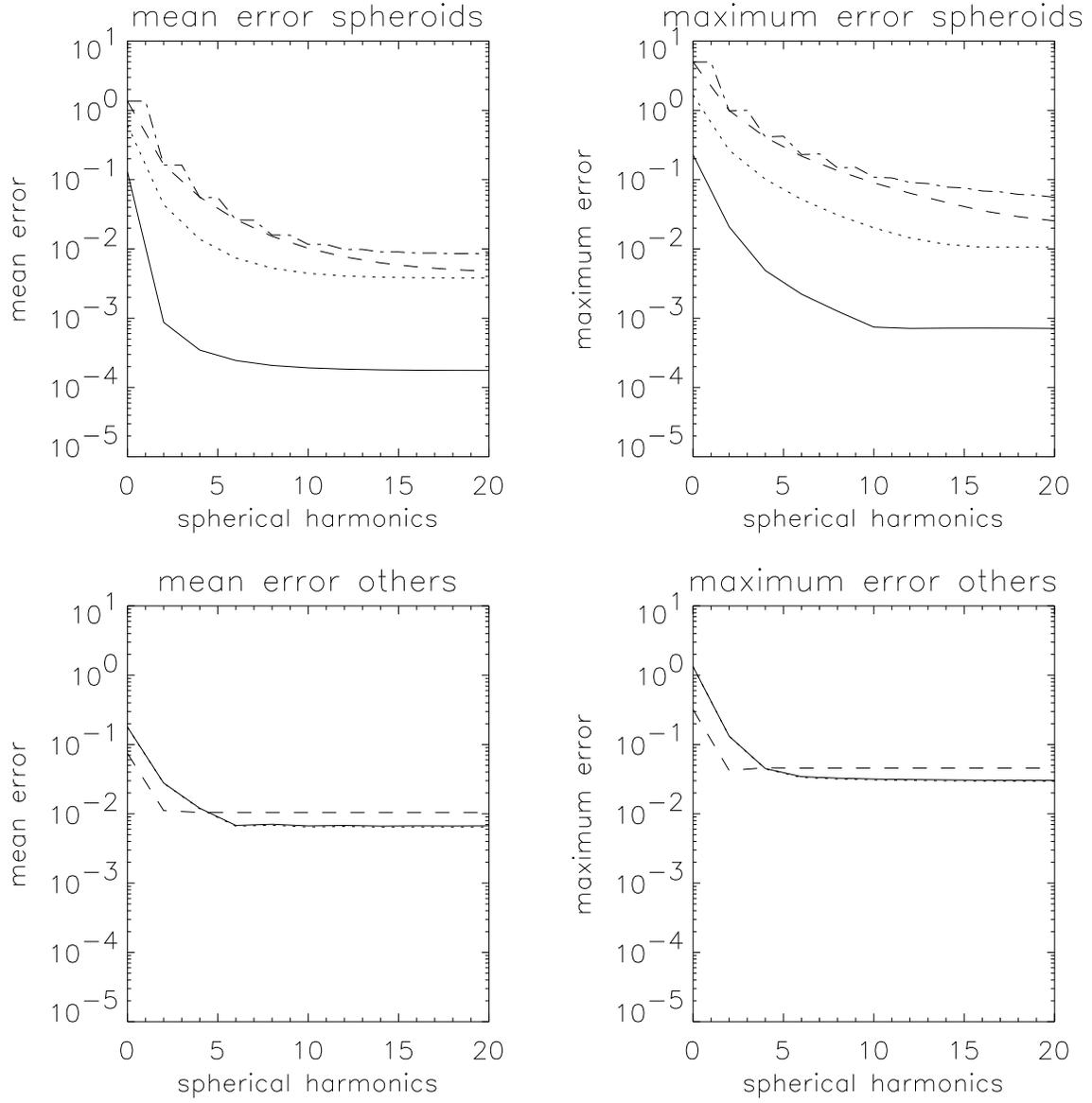

Figure 2: Mean (left) and maximum (right) errors of the potential solver compared to the analytical solution for a grid of 500 by 250 zones. Upper two panels: Errors for the interior solution of a homogenous oblate spheroid with an axis ratio 0.7 (solid) and 0.2 (dotted), respectively. The dashed dotted line corresponds to an oblate spheroid with an axis ratio 0.2, which was computed without assuming equatorial symmetry, while the dashed line shows a prolate spheroid with an axis ratio 0.2. Lower two panels: Errors for the Kuzmin-Plummer disk (solid) and the logarithmic potential (dashed) of Fig. 1, and for a Satoh density. distribution, which for the choice of parameters is similar to the Kuzmin-Plummer disk.

is anyway poorly described on a spherical grid. In summary, the potential solver works well, if the use of spherical coordinates is appropriate

Judging from our experience, in all adequate cases the potential derived from the monopole term alone, is already accurate to within 50%. Including the quadrupole, a mean error of a few percent and a maximum error of about 10% can be achieved in most cases. Using even higher moments, a mean accuracy of the order of a few $10^{-3}$ and a maximum error of about one percent can typically be obtained with $L = 10 \ldots 14$. A further increase of the number of multipole moments does not yield an increased accuracy (see Fig. 2). We find that the errors in the numerical solution are not randomly distributed on the grid, but instead show a smooth overall trend, i.e., the force can be determined with almost the same accuracy as the potential. We encourage the reader to use the program on the enclosed floppy disk, to study the convergence behaviour of the numerical solution for various configurations, grids and number of spherical harmonics.

The computational requirements of the method can be illustrated by the following numbers. A 2D potential calculation on a $500 \times 250$ grid including multipole moments up to $L = 20$ typically requires 0.3 s on a IBM RISC/580 and 0.05 s on a CRAY-YMP. For a 3D grid of $360 \times 18 \times 72$ zones and $L = 14$ the solution is obtained in 1.8 s on a CRAY-YMP assuming equatorial symmetry (Zwerger; private communication). These timings are, therefore, comparable to the requirements of potential solvers based on the FFT technique, which however are much less flexible and from the geometrical point of view (cubic grid, periodic boundaries) less suitable for many astrophysical problems. On the other hand an ADI solver, which iteratively solves Poisson's equation in differential form (see above; Eq. (1)) requires about 50 times more computer time. Thus, while in hydrodynamical simulation the solution of Poisson's equation by the ADI method amounts to a sizeable fraction of the computing time of one hydrodynamical timestep, the fraction is practically negligible for the proposed Poisson solver. Finally, we want to pint out that although the proposed Poisson solver is based on spherical coordinates, it can easily be applied in simulations using Cartesian or cylindrical coordinates. In that case, the density has to be interpolated to an auxiliary spherical grid and after the potential has been obtained, $\Phi$ has to be interpolated back to the original grid. Test calculations show, however, that the computing time for the two interpolations can exceed the time for actually solving Poisson's equation.

## 3. Implementation of a Poisson Solver into a Hydrodynamical Code

In order to implement the above described (or any other) Poisson Solver into a hydrodynamical code one has to proceed with some care, because otherwise pitfalls, like e.g., a violation of the total energy conservation may occur when calculating self-gravitating flows. We begin by presenting the hydrodynamical equations in conservation form and then outline their numerical solution by means of so-called Godunov-type methods. This class of finite difference methods, and in particular one of its higher-order versions PPM (piecewise parabolic method; see below) has become increasingly popular in astrophysics in recent years. Thus, we only consider these methods here, and also restrict the subsequent discussion of the implementation of the Poisson solver to this class of methods. However, we point out that most of the considerations also hold for any other conservative finite difference method.

### 3.1 The Hydrodynamical Equations of Self-gravitating Flows

In the absence of viscosity and for the special case of an *adiabatic flow* the hydrodynamical equations of self-gravitating flows can be written in Eulerian form as (see e.g., Landau &

Lifshitz 1959)

$$\frac{\partial \varrho}{\partial t} + \text{div}(\varrho \mathbf{v}) = 0 \,. \tag{21}$$

$$\frac{\partial \varrho \mathbf{v}}{\partial t} + \text{div}(\varrho \mathbf{v} \otimes \mathbf{v}) + \text{grad} P = -\varrho \text{grad} \Phi \,, \tag{22}$$

$$\frac{\partial \varrho E}{\partial t} + \text{div}\left[(\varrho E + P)\mathbf{v}\right] = -\varrho \mathbf{v} \text{grad} \Phi \,, \tag{23}$$

where $\rho$ is the fluid density, $\mathbf{v}$ is the fluid velocity, $P$ is the isotropic fluid (or gas) pressure, $E = \frac{1}{2}|\mathbf{v}|^2 + \varepsilon$ is the total energy per unit mass (erg/g), $\varepsilon$ is the internal energy per unit mass (erg/g), and $t$ is the time coordinate. The pressure is related to the energy and the density by the equation of state, which for an adiabatic flow is given by a simple gamma-law relation

$$P = (\gamma - 1)\varrho\varepsilon \,, \tag{24}$$

where $\gamma$ is the ratio of specific heats.

## 3.2 Godunov-type Methods for Solving the Hydrodynamical Equations

One of the most successful areas of research in recent years for improving the accuracy of numerical methods for hydrodynamics has been in the class of schemes known as high-order Godunov methods (for an introduction see e.g., Oran & Boris 1987, LeVeque 1992). In these methods, the flow is divided into a series of slabs, each of which occupies a single zone of the computational grid. The discontinuities between these slabs are treated by solving Riemann's shock tube problem at each zone interface. This has the effect of introducing non-linearity explicitly into the finite difference equations and allows the method to calculate sharp shock fronts and contact discontinuities without introducing significant oscillations. Usually, discontinuities can be confined to only one to two zones in width.

The original method of Godunov (1959) approximated the flow variables by a piecewise-constant distribution, resulting in a scheme which is accurate to only first order. A significant advance was made in the MUSCL scheme (van Leer 1979), in which second-order accuracy is achieved by representing the flow variables by piecewise-linear functions. This is analogous to switching from the rectangle rule to the trapezoidal rule for integrating a function. Another major advance in this scheme was the use monotonicity constrains to eliminate oscillations in the flow. Later on Colella & Woodward (1984) proposed the Piecewise Parabolic Method (PPM), which achieves even higher accuracy by representing the flow variables in each zone by monotonized parabolas.

The first step in the calculation is to construct values at each zone interface for $\rho$, $u$, and $P$ from the zone-average values. The zone-average value of $P$ is obtained from $\rho$, $u$, and $E$ using the equation of state. This approximation is accurate to second order. It is not appropriate to interpolate the interface values by passing a polynomial through the values of $\langle\rho\rangle$, $\langle u\rangle$, and $\langle P\rangle$, since these are average values rather than point values. A more complicated algorithm is required, which involves constructing an interpolating polynomial which has the correct average value in each zone. A cubic interpolation polynomial, which involves the average values in four zones, is used. The exact method by which this formula is derived is explained in Colella & Woodward (1984).

The careful treatment of the discontinuities which occur at the zone interfaces is one of the most important aspects of PPM (as well as all Godunov schemes). The procedure which is used involves the solution of Riemann's shock tube problem at each zone interface. Riemann's problem can be stated very simply. A discontinuity separated by two constant states will, in general, produce three hydrodynamic waves. Two non-linear waves (shocks or rarefactions) will propagate in opposite directions. A contact discontinuity (i.e., a jump in density with continuous pressure and velocity) will remain in the middle. In Lagrangian coordinates, the contact discontinuity remains at the zone interface. These three waves will be separated by states having values which remain constant in time. The complete solution of Riemann's problem involves calculating the evolution of these three waves.

In the PPM method, the two states separated by each discontinuity have spatial distributions which are parabolic rather than constant. Therefore, before solving Riemann's problem, it is necessary to determine what should be used in place of the constant left and right states. In an exact solution of this problem (if it could be obtained), the three waves would no longer be separated by constant states, and a time-dependent solution would have to be obtained.

However, a much simpler procedure produces acceptable results. A spatial average over the parabolic distribution is used in place of the "constant" state. Rather than using a spatial average over the entire zone, the average is computed over only that portion of the zone which can influence the zone interface during the time step. This region is determined by tracing characteristics, i.e., the paths in the $r-t$ plane along which sound waves propagate (assuming a constant sound speed along the characteristic). The two regions which can influence the zone interface during the time step are the intervals between the zone interface and the points of intersection of the characteristics with the horizontal axis. The two constant states are then approximated by the average values of the variables over these two intervals.

The above procedure must be modified slightly if there is an external force, such as gravity. Instead of solving Riemann's problem with the corresponding source terms included, the left and right states are altered so that the resulting solution will be the same as the solution to the full Riemann problem including the source terms to second order accuracy.

Extending a Godunov-type scheme to two (or three) dimensions is not complicated. The method used is often called operator splitting or dimensional splitting (see e.g., Yanenko 1971; Oran & Boris 1987), in which each time step consists of a one-dimensional calculation along one coordinate direction of the grid followed by a one-dimensional calculation along the orthogonal coordinate direction (or the two orthogonal directions in 3D problems). Each one-dimensional sweep accounts for the gradients only in the direction of that sweep. By combining the two (or three) sets of sweeps, all of the terms in the hydrodynamic equations are included. In order to preserve second order accuracy, the order of the one-dimensional calculations must be alternated between time steps.

### 3.3 Incorporating Self-gravity

As mentioned above in conservative finite difference schemes any state variable $U_{ijk}^n$ (e.g., density, energy and velocity) which is assigned to a zone $(i,j,k)$ of volume $\Delta V_{ijk}$ at time $t=t^n$ should be interpreted as the corresponding zone average, i.e.,

$$U_{ijk}^n = \int_{\Delta V_{ijk}} dV \, U(\mathbf{r}, t^n). \tag{25}$$

The question of how to include non-local source terms, like e.g., gravity, in such a method is by no means trivial and conceptional problematic. Because the exact implementation depends on the specific numerical scheme, we restrict ourselves to some more general considerations.

We propose to treat self-gravity in the numerical solution of the hydrodynamical equations by means of an operator-splitting technique. In each timestep one first calculates the change of all state variables due to the advection terms, i.e., due to the divergence terms in the hydrodynamical equations. With the updated density one then solves Poisson's equation to obtain the new potential. Using time centered values of the potential gradient one finally calculates the change of velocity (or momentum) and energy, which completes the work for a timestep. Whether such an operator-splitting approach actually works is not at all obvious, because in many self-gravitating flows the pressure and potential gradients are large and the corresponding forces almost cancel each other. However, experience shows that the operator-splitting approach indeed works and can be used to attack various astrophysical problems of self-gravitating flows (see e.g., Müller, Mair & Hillebrandt 1988; Ruffert & Müller 1992; Steinmetz, Müller & Hillebrandt 1992; Müller 1993).

Let us now consider the implementation of self-gravity in somewhat more detail. For this purpose we restrict ourselves to one-dimensional problems, which significantly simplify the notation but still illustrate the basic ideas. The zone average of the velocity $v(x,t)$ of a zone $i$ at time $t^n$ is given by (see Eq. (25))

$$v_i^n = \frac{1}{x_i - x_{i-1}} \int_{x_{i-1}}^{x_i} dx\, v(x, t^n), \tag{26}$$

where $x_{i-1}$ and $x_i$ are the coordinates of the zone interfaces in negative and positive x-direction, respectively.

For an Eulerian grid (fixed coordinates) the upper and lower limit of the integral in the above equation are time independent. Consequently, one can write the temporal change of the zone averaged velocity due to the action of gravity as

$$\frac{d}{dt} v_i^n = \frac{d}{dt} \left\{ \frac{1}{x_i - x_{i-1}} \int_{x_{i-1}}^{x_i} dx\, v(x, t^n) \right\} \tag{27}$$

$$= \frac{1}{x_i - x_{i-1}} \int_{x_{i-1}}^{x_i} dx\, \frac{dv(x, t^n)}{dt} \tag{28}$$

$$= -\frac{1}{x_i - x_{i-1}} \int_{x_{i-1}}^{x_i} dx\, \frac{d\Phi(x, t^n)}{dx} \tag{29}$$

$$= -\frac{1}{x_i - x_{i-1}} \int_{x_{i-1}}^{x_i} d\Phi(x, t^n). \tag{30}$$

The last equation can be integrated formally, which gives

$$\frac{d}{dt} v_i^n = \frac{\Phi_i^n - \Phi_{i-1}^n}{x_i - x_{i-1}} \tag{31}$$

Obviously, it is advantageous to define the potential at zone interfaces, because then Eq. (31) is a second-order (spatially) accurate approximation of the acceleration of the fluid in zone $i$ at time $t^n$ due to gravity.

Up to now we have implicitly assumed that $x$ is a Cartesian coordinate. If one deals with problems in spherical geometry, one has to treat an integral of the form

$$\frac{3}{r_i^3 - r_{i-1}^3} \int_{r_{i-1}}^{r_i} dr\, r^2\, d\Phi(r, t^n) \tag{32}$$

Contrary to the Cartesian case the integral in Eq. (32) cannot formally be integrated. However, except for the very center of the spherical grid the difference between the Cartesian and the spherical expression is small. Thus, according to our experience it is sufficient to use Eq (31) even for spherical problems. If, however, self-gravitating flows are simulated which depend critically on an accurate treatment close to the coordinate center (e.g., gravitational collapse; see Janka, Zwerger & Mönchmeyer 1993) we recommend a more accurate approach (see also Mönchmeyer & Müller 1989). Actually, by a Taylor series expansion of the integrand in Eq. (32) around the center of mass $r_c$ of the zone, one can show that

$$\int_{r_{i-1}}^{r_i} dr\, r^2\, d\Phi(r, t^n) = \left.\frac{d\Phi}{dr}\right|_{r_c} + \mathcal{O}(\Delta r^2) \tag{33}$$

Since up to second order accuracy the mass and volume center of a zone are equivalent, in many problems one can replace $r_c$ by the volume center $(r_{i-1} + r_i)/2$ without significantly degrading the numerical approximation of the gravitational acceleration term.

In case of a Godunov type scheme, like e.g., PPM, it is important to incorporate the potential in the Riemann solver, too. Similarly to the treatment of fictitious forces or of a constant external force in PPM (see Colella & Woodward 1984), the time centered left and right velocity state of the Riemann problem at the zone interface $i + 1/2$ should be modified according to

$$u_{i,R}^{n+1/2} \rightarrow u_{i,R}^{n+1/2} - \frac{\Delta t}{2} \left(\nabla \Phi_i^{n+1/2}\right)_R \tag{34}$$

$$u_{i,L}^{n+1/2} \rightarrow u_{i,L}^{n+1/2} - \frac{\Delta t}{2} \left(\nabla \Phi_i^{n+1/2}\right)_L, \tag{35}$$

where $\Delta t = t^{n+1} - t^n$. This modification of the state vector corresponds to a time centered prediction of the gravitational force. In principle, $(\nabla \Phi_i^{n+1/2})_{R,L}$ should be the average of the time centered potential gradient over the region which can influence the cell interface within a timestep (see previous subsection). Since the potential is a spatially quite smooth quantity which in addition also changes relatively slowly with time, it is sufficient to replace $(\nabla \Phi_i^{n+1/2})_{R,L}$ by $\nabla \Phi_i^n$, i.e., by the potential gradient at the zone interface at time $t^n$. Note, however, that some extra care must be taken into account if this approach is generalized to moving grids.

Having solved the Riemann problems with the modified left and right state vectors at each interface one can calculate the corresponding flux terms and update the state vectors. After this advection step, which also includes the effects of pressure and fictitious forces, the velocity and the (total) energy are corrected via

$$\mathbf{v}_i^{n+1} = \widetilde{\mathbf{v}}_i^{n+1} - \frac{\Delta t}{2} \left(\nabla \Phi_i^n + \nabla \widetilde{\Phi}_i^{n+1}\right) \tag{36}$$

$$E_i^{n+1} = \widetilde{E}_i^{n+1} - \frac{\Delta t}{2} \left(\mathbf{v}_i^n \nabla \Phi_i^n + \mathbf{v}_i^{n+1} \nabla \widetilde{\Phi}_i^{n+1}\right), \tag{37}$$

which completes the hydrodynamical timestep. The quantities $\widetilde{\mathbf{v}}_i^{n+1}$, $\widetilde{E}_i^{n+1}$ and $\widetilde{\Phi}_i^{n+1}$ denote the velocity, the total energy (per unit mass) and the potential after the advection step. Concerning the discretized Eqs. (36) and (37) we would like to point out an important empirical finding. We have experimented with several other possible forms of discretization (i.e., different from that given in Eqs. (36) and (37)), but all of them led to inferior results and some even to an abort of the code caused by a severe violation of the total energy conservation.

## 4. Conclusions

We have presented an efficient solver for Poisson's equation which can be used in hydrodynamic calculations. The proposed solver is based on an expansion of the integral form of Poisson's equation into spherical harmonics and thus naturally utilizes spherical coordinates. The resulting algorithm has a computational cost being proportional to $(L + 1)^2 \times N$, where $L$ and $N$ are the number of spherical harmonics used in the expansion and the number of grid zones, respectively. The solver can easily be programmed as is illustrated by the 2D version of the algorithm delivered with this volume on a floppy disk. The solver can be applied to and works well for all problems for which the use of spherical coordinates is appropriate. For very flat disk-like configurations or for problems which involve two or more spatially separated distinct mass concentrations (e.g., a binary) the proposed algorithm should thus not be applied.

The implementation of the solver into modern Godunov-type hydrodynamic codes needs some careful considerations in order to avoid large errors in the total energy conservation. Practically, the inclusion of self-gravity requires an additional source term in the momentum and energy equation, and a simple modification of the velocity state defining the local Riemann problems. Our experience shows that the resulting hydrodynamic code can be used to solve a large set of interesting self-gravitating flows in various areas of astrophysics.

*Acknowledgements:* The authors would like to thank Bruce Fryxell, Ralph Mönchmeyer, Maximilian Ruffert and Thomas Zwerger for helpful discussions.